\documentclass{eas}
\usepackage{graphicx}

\TitreGlobal{The Title of this Volume}
\begin{document}

\title{Super-massive stars: Radiative transfer} 
\author{Pau Amaro-Seoane, Rainer Spurzem and Andreas Just}
\address{Astronomisches Rechen-Institut, 
Moenchhofstra\ss e 12-14 D-69120 Heidelberg, Germany}
\begin{abstract}
The concept of central 
super-massive stars (${\cal M} \ge 5 \times 10^4 M_{\odot}$,
where ${\cal M}$ is the mass of the super-massive star)
embedded in dense stellar systems was suggested as a possible explanation for high-
energy emissions phenomena occurring in active galactic nuclei and quasars (Vilkoviski 
1976, Hara 1978), such as X-ray emissions (Bahcall and Ostriker, 1975).
SMSs and super-massive black holes are two possibilities 
to explain the nature of super-massive central objects, and super-massive stars may be 
an intermediate step towards the formation of super-massive black holes (Rees 1984).
Therefore it is important to study such a dense gas-star system in detail. We address
here the implementation of radiative transfer in a model which was presented in
former work (Amaro-Seoane and Spurzem 2001, Amaro-Seoane et al. 2002). In this sense,
we extend here and improve the work done by Langbein et al. (1990) by describing the
radiative transfer in super-massive stars using previous work on this subject (Castor 1972).
\end{abstract}
\maketitle
\section{Introduction}
The radiation intensity $I_{\nu}(\theta, \phi)$ is defined as the amount of energy that
passes through a surface normal to the direction ($\theta$,$\phi$) per unit solid angle
(1 steradian) and unit frequency range (1 Hz) in one second. The intensity of the total
radiation is given by integrating over all frequencies.
The three radiation moments (the moments of order zero, one and two) are defined by:
\begin{eqnarray}
&J=&\int_{0}^{\infty} J_{\nu} d{\nu}=\int_{0}^{\infty} d{\nu} \frac{1}{2} \int_{-1}^{+1} 
I_{\nu} d{\mu} \nonumber \\
&
H=&\int_{0}^{\infty} H_{\nu} d{\nu}=\int_{0}^{\infty} d{\nu} \frac{1}{2} \int_{-1}^{+1} 
I_{\nu} \mu d{\mu}  \\
&K=&\int_{0}^{\infty} K_{\nu} d{\nu}=\int_{0}^{\infty} d{\nu} \frac{1}{2} \int_{-1}^{+1} 
I_{\nu} \mu^2 d{\mu}, \nonumber 
\end{eqnarray}
\noindent
The moment of order zero is related to the density of energy of the field of radiation 
($E_{\rm rad}=4\pi J/c$), the moment of order one to the flux of radiation 
($F_{\rm rad}=4 \pi H$) and the moment of order one to the radiation pressure 
($p_{\rm rad}=4\pi K/c$).

The transfer of
radiation in a spherically symmetric moving medium is considered
taking into account the contributions which are of the order of the flow velocity
divided by the velocity of light; we include also the variation from the centre up 
to the atmosphere of the Eddington factor $f_{\rm Edd}=K/J$, where $K$ and $J$ are 
the radiation moments; $f_{\rm Edd}$ is obtained from a numerical solution of the
equation of radiation transfer in spherical geometry (Yorke 1980). 
We get the radiation transfer equations by re-writing the frequency-integrated moment 
equations from Castor (1972) in logarithmic variables in order to study the dense gas-star
system at long-term,

\begin{eqnarray}
{\frac{1}{c} \frac{\partial J}{\partial t}+ \frac{\partial H}{\partial r}+
\frac{2 H}{r}- \frac{J(3f_{\rm Edd}-1)}{cr} u_{g}- \frac {J(1+f_{\rm Edd})}{c}
\frac {\partial \ln \rho_{g}}{\partial t}=} \nonumber \\
\kappa_{\rm abs}(B-J)
\end{eqnarray}

\begin{eqnarray}
\frac{1}{c} \frac{\partial H}{\partial t}+\frac{\partial (J f_{\rm Edd})}{\partial r}
+ \frac{J(3f_{\rm Edd}-1)}{r}-\frac{2u_{g}}{cr} H-\frac{2}{c}\frac{\partial 
\ln \rho_{g}}{\partial t} H= \nonumber \\
-\kappa_{\rm ext} \rho_{g} H
\end{eqnarray}

\noindent
Where we have substituted $F_{\rm rad}=5p_{g}v_{g}/2$.
In the equations $\kappa_{\rm abs}$ and $\kappa_{\rm ext}$ are the absorption and extinction
coefficients per unit mass 
\begin{equation}
\kappa_{\rm abs}=\frac{\rho_{g} \Lambda(T)}{B},~
\kappa_{\rm ext}=\rho_{g}(\kappa_{\rm abs}+\kappa_{\rm scatt}),
\end{equation}
\noindent
$\Lambda(T)$ is the cooling function, $B$ the Planck function and $\kappa_{\rm scatt}$ the
scattering coefficient per unit mass. We have made use of $\partial M_{r}/ \partial 
r=4 \pi^2 \rho$, $f_{\rm Edd}=K/J$, 
and the Kirchhoff's law, $B_{\nu}=j_{\nu}/\kappa_{\nu}$ ($j_{\nu}$ is the emission 
coefficient), so that the right-hand terms in Castor (1972) are the corresponding given here.


\end{document}